\documentclass[twocolumn,showpacs,preprintnumbers,amsmath,amssymb]{revtex4}

\usepackage{graphicx}
\usepackage{dcolumn}
\usepackage{bm}

\begin{document}


\title{Electronic structure of S-C$_6$H$_5$ self-assembled monolayers on
       Cu(111) and Au(111) substrates.}

\author{Vasili Perebeinos$^1$\cite{byline} and Marshall Newton$^2$}
\affiliation{$^1$Department of Physics,\\ $^2$Department of Chemistry,\\
Brookhaven National Laboratory, Upton, New York 1973-5000 }

\date{\today}

\begin{abstract}

We use first principles density functional theory to calculate
the electronic structure of the phenylthiolate (S-C$_6$H$_5$)
self-assembled monolayer (SAM) on Cu(111) and Au(111) substrates.
We find significant lateral dispersion of the SAM molecular states
and discuss its implications for transport properties of the
molecular wire array. We calculate the two photon photoemission
spectra and the work function of the SAM on Cu(111) and compare them
with the available experimental data. Our results are used to
discuss assignments of the observed spectral data and yield
predictions for new electronic states due to the monolayer not yet
accessed experimentally.
\end{abstract}

\maketitle

\section{\label{sec1}Introduction}

Recently there has been increased interest in the possibility of
using organic molecules as electronic components in nanoscale
devices.  One of the most important issues is the role of
molecule-metal contact. The charge injection at the metal-molecule
interface is usually modelled by the energy dependent probability for a
substrate electron to tunnel to the molecular wire. The
magnitude of the tunnelling probability depends on lineup of the
molecular wire electronic states and the metal Fermi level. When a
closely packed array of molecular wires is formed on the metal
electrode, the molecular states form bands dispersive in the
lateral direction. This suggests that the tunnelling probability
in addition to the energy of the injected electron depends also on
the lateral component of the electron wavevector.

The change of the metal work function due to a self-assembled
Monolayer  (SAM) may be employed in   device applications to
assist the charge injection into the molecular wire
\cite{Phaedon}. The work function of a crystal surface has
generally two contributions \cite{Lang}: (1) the electrostatic
barrier due to the distortion in the charge distribution at the
surface and (2) the many-body effects of the screened hole.
It is important to know the relative contributions of the two
components for the successful design of   molecular electronic
devices.

In this paper we present Density Functional Theory (DFT)
calculations of the electronic structure of  SAMs of phenylthiolate
(PT = -S-C$_6$H$_5$)  on noble metal substrates Cu(111) and
Au(111). We find significant lateral dispersion of the molecular
electronic states derived from the molecular orbitals of the SAM.
Our calculations aim to address three essential questions: (1)
what are the molecular states closest to metal Fermi level and
therefore the most important for transport properties; (2) how
large is the lateral dispersion of those states; (3) what is the
largest contribution to the workfunction change associated with SAM formation,
for which we find the electrostatic contribution from the polar molecules as
determined by the dipole moment of the isolated array of molecules in addition
to that of the bare metal surface. In addition
we calculate two-photon photoemission (2PPE) spectra using DFT
wavefunctions and the work function of the SAM/Cu(111) system to
compare with the experiments by Zhu {\it et. al.} \cite{Zhu}.

\section{\label{sec2}Method of calculation}

We use the full potential linearized augmented plane wave
(FP-LAPW) method \cite{Singh1} with local orbital extensions
\cite{Singh2} in the WIEN2k implementation \cite{Blaha}. The GGA
\cite{Perdew} exchange-correlation potential was used.
Well-converged basis sets were employed with a 4.7 Ry plane wave
cut off. Five special k-points were used to sample the two
dimensional Brillouin zone. The surface plane was taken as the $x,
y$ plane. We used slab calculations based on a super cell with
three metal layers covered on one side by a PT monolayer for
electronic structure calculations, and six metal layers with both
sides covered for work function calculations. The PT monolayer is
of the so called $\sqrt{3}\times\sqrt{3}R30^0$ type
\cite{Schreiber}, with one molecule per three substrate surface
metal atoms. The former are arranged in a triangular lattice with
side $\sqrt{3}R_{M-M}$, where $R_{M-M}$ is the side of the
triangular lattice formed by metal atoms $M$ in the (111) plane. For $M$=Au
and Cu, the corresponding cross-section area for molecule is,
respectively, 21.6 \AA$^2$ and 16.9 \AA$^2$
\cite{Schreiber,Leung}. The cross-sectional mean area of a PT
molecule (when directed normal to the substrate) is 21.1 \AA$^2$,
slightly less than the above value based on the Au substrate, thus
indicating the possibility of a small degree of tilt
\cite{Schreiber,Leung}, whereas no tilt is expected in the case of
the Cu substrate.

The geometry of the molecular array has been fully relaxed on an
unrelaxed substrate with metal atoms in ideal crystal positions.
The monolayer on the Au substrate was found tilted by 18$^0$,
consistent with related experimental data \cite{Leung}. In the
calculation, the PT molecule was constrained so as to maintain
$C_{2v}$ intramolecular symmetry \cite{geometry}. The
lowest-energy binding site was the Au-Au bridge, analogous to the case
of alkanethiols studied by Selloni {\it et. al.} \cite{Selloni}.
The S atom height is 2.1 \AA . \ Starting with alignment normal to
the substrate, with the projection of the PT plane parallel to a
nearest-neighbor Au-Au vector, the PT was tilted in the direction
perpendicular to its plane (tilting in the orthogonal direction
caused a sharp rise in energy). The calculated tilt relaxation
energy per molecule ($0^0$ to $18^0$) was of the order of $k_BT$
at room temperature, indicating the likelihood of appreciable
thermal fluctuations, similar to the situation noted in
\cite{Schreiber,Leung}.

In the case of the Cu substrate we find zero tilt angle for the
monolayer due to the close spacing controlled by the Cu lattice
constant, with the lowest energy when sulfur binds to the hollow
site. The height of the S atom is 1.73 \AA.\ \ The binding energy
for the bridge site is higher by 170 meV. The frozen lattice
approximation used here may cause a shift of the energy difference
in binding energy of a few tens of meV for different adsorption
sites, but it should not affect much the work function and the
shape of the dispersion curves.

\section{\label{sec3}Band structure results}

\begin{figure}
\includegraphics[height=2.36in,width=3.16in,angle=0]{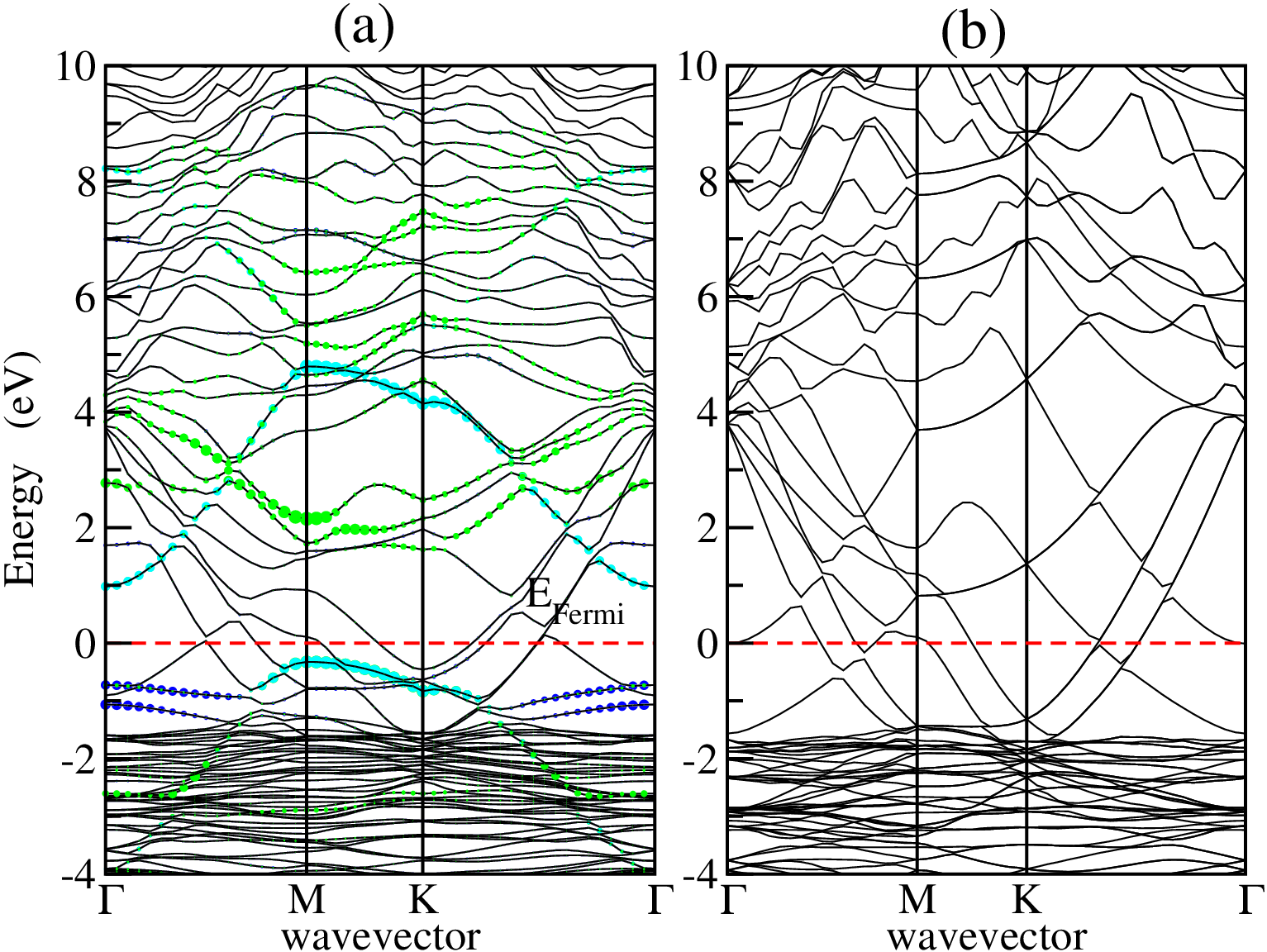}
\caption{\label{fig2} (color online). Band structure results
(energies relative to E$_{\rm Fermi}$). (a) PT SAM on 3 layers of
Cu(111). The size of the colored circles is proportional to the
projected density of states of each band on a given atom of a
molecule: sulfur - blue, C$_{\rm s}$ carbons - light blue, and
C$_{\rm t}$ carbons - green. (b) 3 layer slab of Cu(111). From
comparison of (a) and (b) four HOMO and three LUMO bands can be
identified  due to the SAM attachment.}
\end{figure}

The electronic structure along the high symmetry direction of the
2D Brillouin zone is shown in Fig. (\ref{fig2}a) for the PT SAM on
Cu(111). In order to distinguish which bands originate from the
SAM and which are due to the metallic substrate we calculate
projected density of states of each band inside the spheres
surrounding the atoms of the PT molecule \cite{Singh1,Singh2}. The
Cu surface states were identified by the similarities of the band
dispersion of the covered and clean Cu(111) three layer substrates
shown on Fig. (\ref{fig2}b). At least seven molecular bands due to
the SAM can be identified. These molecular bands can be
characterized by (1) the energy at the $\Gamma$-point, (2) the
effective mass for the $\Gamma$-M direction in cases where a
parabolic fit of the dispersion in the vicinity of the $\Gamma$
point is possible, (3) the bandwidth defined as the energy
difference between the M and the $\Gamma$ point, and (4) the major
atomic contributions. These band characteristics are summarized in
Table \ref{tab1}.

\begin{table}
\begin{ruledtabular}
\begin{tabular}{crrrrrrc}
band & \multicolumn{2}{c}{E$_{\Gamma}$ (eV)} &
\multicolumn{2}{c}{m$_{eff}$ $(m_e)$} &
 \multicolumn{2}{c}{E$_M$-E$_{\Gamma}$ (eV)} &  Character \\
                & SAM/Cu & SAM     &SAM/Cu & SAM   & SAM/Cu & SAM & \\
\hline
1           &  -4.0  &  -3.6   &       &       & 3.7 & 3.86 & C$_{\rm s}$\\
2           &  -2.7  &  -1.9   &       & -1.5  &     &      & C$_{\rm t}$\\
3           &  -1.1  &  0.30   & -2.0  & -1.0  &     &      & S\\
4           &  -0.75 &  0.35   & -2.0  & -1.2  &     &-0.88 & S\\
5           &   1.0  &   1.6   &  0.4  & 0.35  & 3.8 & 3.8  & C$_{\rm s}$\\
6           &   2.8  &   3.7   &  -0.4 &  1.0  &     &      & C$_{\rm t}$\\
7           &   4.0  &   5.0   &  -0.3 & -0.8  &-1.8 &-2.0  & C$_{\rm t}$\\
SS$_{\rm o}$  &  -0.9  &         & 0.4   &       &     &      & Cu \\
SS$_{\rm u}$          &   1.7  &         &       &       &     &      & Cu \\
\hline
                & SAM/Au& SAM    &SAM/Au & SAM  &SAM/Au &SAM & \\
1           & -2.4  & -2.2   & 0.55  &  0.8 & 1.6  & 1.5 & C$_{\rm s}$  \\
2           & -2.4  & -2.1   &       &      &      &-0.2 & C$_{\rm t}$ \\
3           & -0.98 & 0.13   & -2.3  & -2.7 &      &-0.2 & S \\
4           & -0.85 & 0.09   & 1.6   & -3.9 &      &-0.3 & S \\
5           &  2.75 & 2.9    &  1.0  & 0.95 &  1.2 & 1.1 & C$_{\rm s}$ \\
6           &  4.1  & 3.9    &       & 0.9  &      &     & C$_{\rm t}$\\
7           &  4.1  & 4.3    & -0.6  &-1.2  &      & -1.0& C$_{\rm t}$\\
 SS$_{\rm o}$ & -0.45 &        &  1.7  &      &      &     & Au \\
 SS$_{\rm u}$         &  1.4  &        &  0.5  &      &      &     & Au \\
\end{tabular}
\end{ruledtabular}
\caption{\label{tab1} Analysis of the electronic bandstructure for
the PT SAM on Cu(111), SAM on Au(111), and the free molecular
arrays at the same geometries. The energies of the bands are given
relative to the Fermi level at the $\Gamma$-point ($k=(0, 0)$).
Effective masses are reported for the bands where a parabolic fit
(along the $\Gamma$-M direction) is possible. The bandwidth is
reported for bands where the energy difference between the M and
the $\Gamma$ point can be identified (M denotes $k=(4\pi/\sqrt{3},
0)$). The last column shows the band character for the
SAM/substrate systems, where C$_{\rm s}$ stands for side phenyl
carbons and C$_{\rm t}$ for next to sulfur and top carbons.}
\end{table}

For the copper substrate the molecular levels closest to the Fermi
level are the two HOMO's (bands 3 and 4), which are of S (3p)
character. They have a very small dispersion (of about 0.2-0.3 eV)
and the parabolic fits result in large effective masses (-2.0
$m_e$). The second occupied state (SS$_{\rm o}$, at -0.9 eV) and
the second unoccupied state (SS$_u$, at 1.7 eV) are surface
states, which are at energies 0.0 eV and -1.6 eV, respectively, on
 the clean substrate in Fig. (\ref{fig2}b).
The discrepancy with the experimental value of the -0.4 eV surface
state energy at $\Gamma$ point is due to the finite size effect of
our three layer slab geometry \cite{Euceda}. The largest lateral
dispersion is due to the bands 1 and 5 originating from the side
phenyl carbons (C$_s$ for short). (The minimal separation between
such atoms on neighboring PT molecules is 2.61 \AA.) The molecular
bands 2, 6, and 7 are mainly due to the next to sulfur and top
carbon atoms (C$_t$ for short). These C$_t$ atoms form the longest
intermolecular contacts and the lateral dispersion of the
C$_t$-type bands is not significant compared to that of side
carbon C$_s$-type bands. There are certainly more electronic
states in the band structure due to the SAM, as is evident from
the band count in Fig. (\ref{fig2}), but these are high energy
states (relative to Fermi level) and are not important for the
conduction properties of the monolayer. These states are also well
delocalized in space such that projected density of states inside
the muffin tin spheres does not give a unique identification of
the molecular bands. As will be clear in Section \ref{sec4}, the
band 6 at 2.8 eV (the $\Gamma$ point) and   one at 6 eV have
the largest contribution to the 2PPE intensity and thus are
candidates for the peaks observed by Zhu et. al. \cite{Zhu}.

\begin{figure}
\includegraphics[height=2.27in,width=3.147in,angle=0]{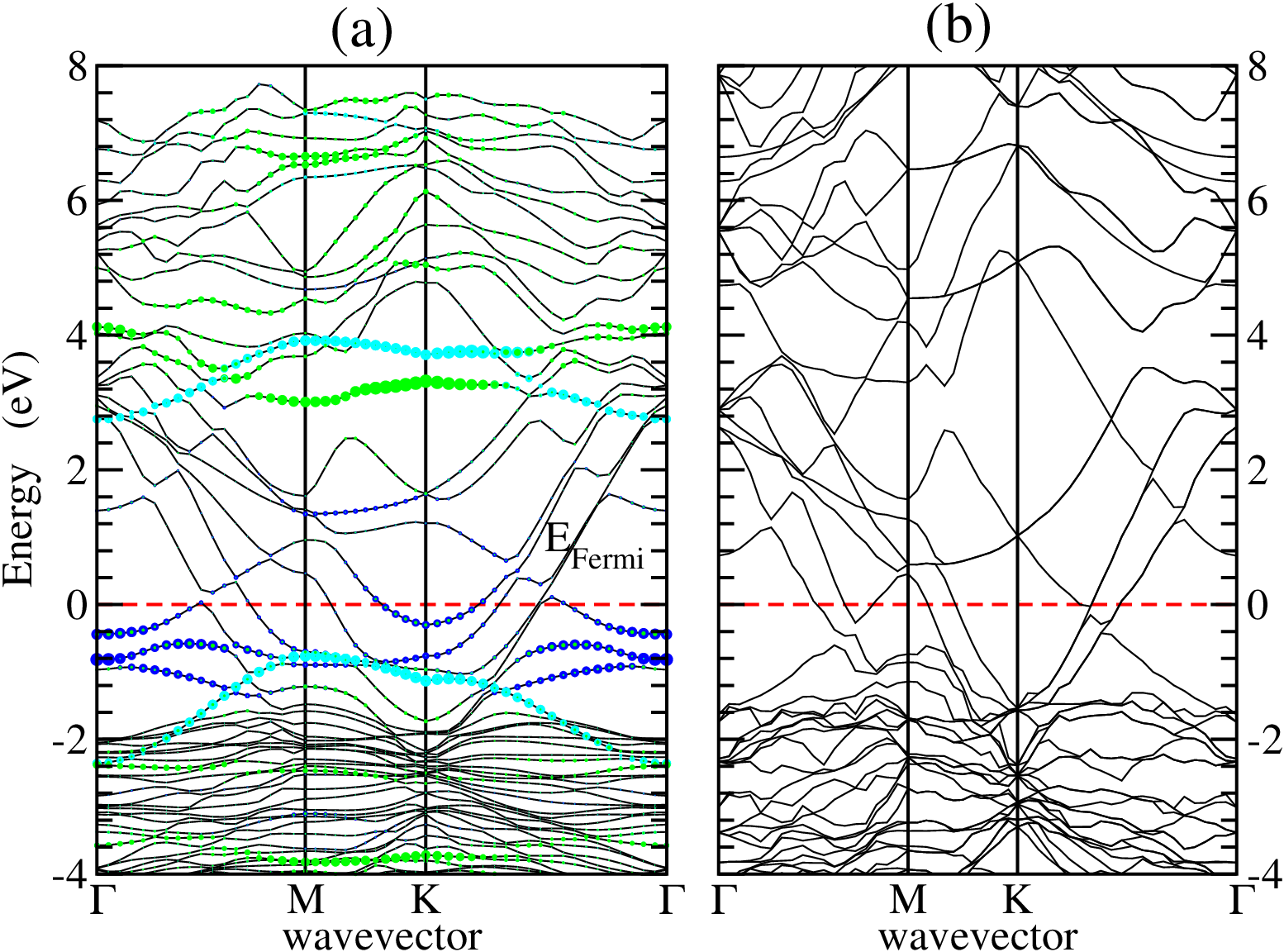}
\caption{\label{fig3} (color online). Band structure results
(energies relative to E$_{\rm Fermi}$) (a) PT SAM on 3 layers of
Au(111). The size of the colored circles is proportional to the
projected density of states of each band on a given atom of a
molecule: sulfur - blue, C$_{\rm s}$ carbons - light blue, and
C$_{\rm t}$ carbons - green. (b) 3 layer slab of Au(111).}
\end{figure}

When the PT is attached to the gold substrate, the bandwidths of
all molecular levels are narrower than for the Cu case, because the nearest
neighbor
distance between the molecules is larger due to the larger lattice
constant of the underlying substrate. The C$_s$ atoms on
neighboring molecules of the SAM (18\% tilted on Au) are separated
by at least 3.13 \AA \ \ from each other, whereas C$_t$ carbons
are 5.0 \AA \ \ apart (i.e., $\sqrt{3}R_{Au-Au}$). The electronic
band structures of the SAM/Au(111) and the clean Au(111) surfaces
are shown in Fig. (\ref{fig3}). The band assignments are
summarized in table (\ref{tab1}). As in the case of the SAM on Cu,
the two nearly degenerate sulfur-based HOMO's (bands 3 and 4) are
the molecular levels closest to the Fermi level. As the lattice
constant increases (in proceeding from Cu to Au), the direct
exchange mechanism responsible for the negative effective mass
decays faster than the substrate mediated mechanism (see below). This causes
the sign of the effective mass of one of the sulfur bands to
become positive on the Au substrate. The most dispersive C$_s$
bands 1 and 5 have bandwidth smaller by a factor of 2.5 than for
the Cu substrate. The occupied surface state SS$_{\rm o}$ is
strongly hybridized with the sulfur 3p orbitals.

\begin{figure}
\includegraphics[height=1.972in,width=3.27in,angle=0]{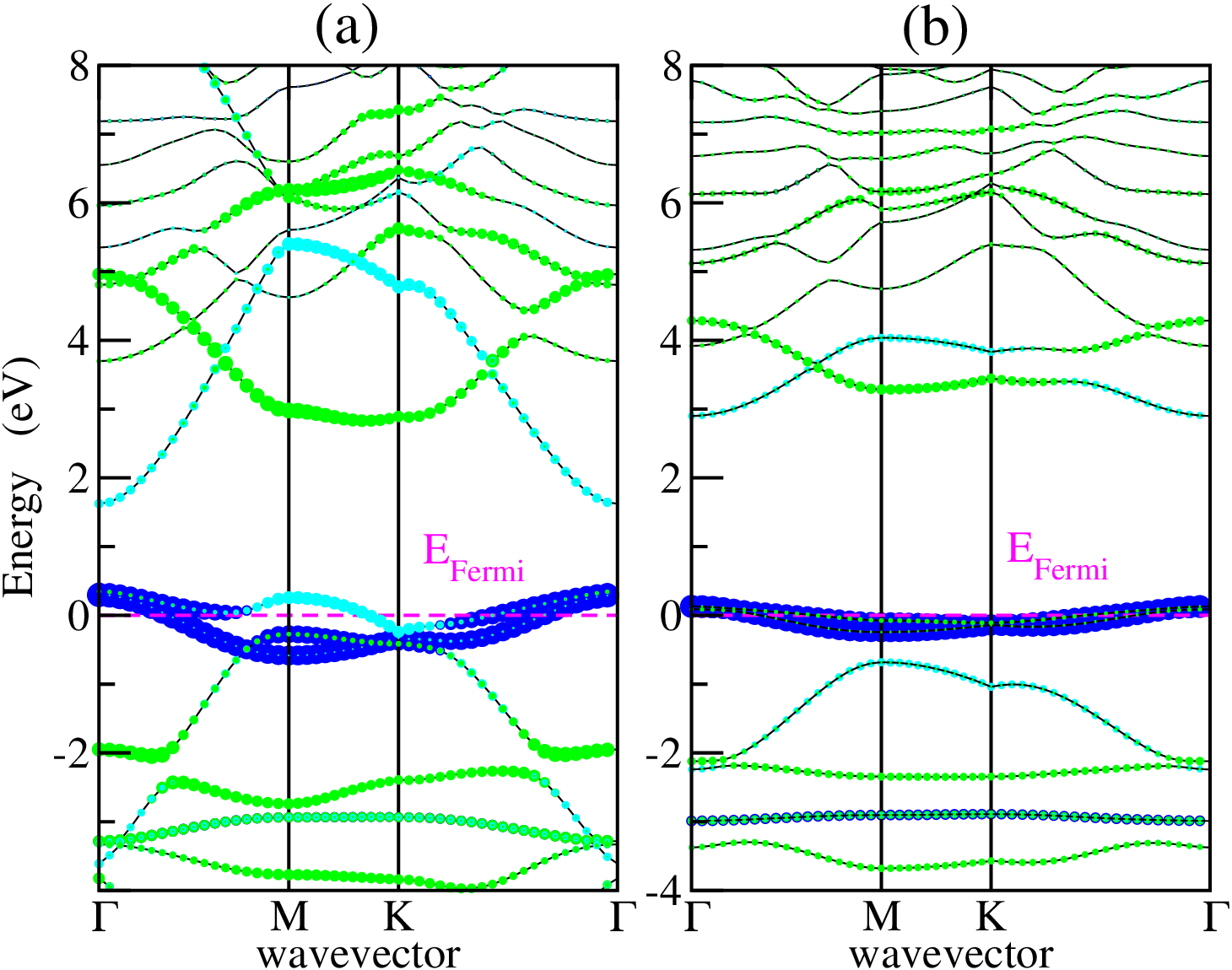}
\caption{\label{fig4} (color online). (a) Band structure of the
free standing  PT array, based on the Cu(111) lattice constant
and (b) a PT array tilted by 18$^0$, based on the Au(111) lattice
constant (energies relative to E$_{\rm Fermi}$). The size of the
colored circles is proportional to the projected density of states
of each band on a given atom of a molecule: sulfur - blue, C$_{\rm
s}$ carbons - light blue, and C$_{\rm t}$ carbons - green.}
\end{figure}

The lateral dispersion has two contributions the direct overlap
between the molecules and the substrate mediated interaction. To
estimate the relative importance of the two mechanisms for the band
dispersion we compare the lateral dispersions obtained for the PT
arrays in the absence of the substrates (infinitely separated from
the substrate), but maintaining the same geometry as determined
with the substrate present. In this case the band structure shown
in Fig. \ref{fig4} has only a direct contribution. The signs of the
dispersions for bands 3 and 4 are negative and the bandwidth is smaller
for the Au(111) case (Fig. (\ref{fig4}b)). In both cases the
dispersion of the sulfur bands is larger in the absence of the
substrate. This suggests that substrate-mediated interactions have
an effect on the bandwidth opposite to that from the direct
exchange. When the lattice constant is relatively large (i.e. the
Au case), the substrate-mediated interaction even changes the sign of
the band 4 dispersion. The coupling of the C$_{\rm s}$ atoms to the substrate
is much smaller, so that substrate-mediated interaction in bands 1
and 5 is responsible for only 10\% of the total bandwidth.

\section{\label{sec4}2PPE spectra}

The electronic structure of the organic SAM can be directly probed
in the 2PPE experiment \cite{Zhu2}. There are three electronic
states involved in the 2PPE spectra: initial ({\it i}),
intermediate ({\it k}), and the final ({\it f}). If one of those
states is localized and has the largest dipole matrix element in comparison
with the rest of the states, then three scenarios are possible for the
variation of kinetic energy ($?E_{kin}$) with respect to the excitation
photon energy $\hbar\omega_L$, which is chosen to be less than the
workfunction of the substrate: (1) $?E_{kin}\approx
2\hbar\omega_L$; (2) $?E_{kin}\approx \hbar\omega_L$, and (3)
invariance with respect to $\hbar\omega_L$. For PT on Cu(111),
two cases are reported \cite{Zhu}: case(2) in which the
intermediate state (i) is assigned as a LUMO lying below the vacuum at 3.3 eV
and case(3), in which the final state (f) is assigned as a
LUMO lying above the vacuum at 6.4 eV
(relative to E$_{\rm Fermi}$). Case (1), corresponding to an initial
HOMO state and was not observed in \cite{Zhu}.

In the present work we use DFT wavefunctions to calculate the 2PPE
spectra at the $\Gamma$-point.  We find two states to contribute most to the
2PPE intensity: a LUMO (band 6) at 2.8 eV (of C$_{\rm t}$ carbon
character) and high energy bands at around 6.0 eV, whose character
is complex due to hybridization of molecular and Cu states. This assignment is
consistent with the conclusion
drawn in \cite{Zhu} based on similarities of spectra taken for SAMs based on
PT and alkanethiolates of different lengths.

\begin{figure}
\includegraphics[height=2.18in,width=2.97in,angle=0]{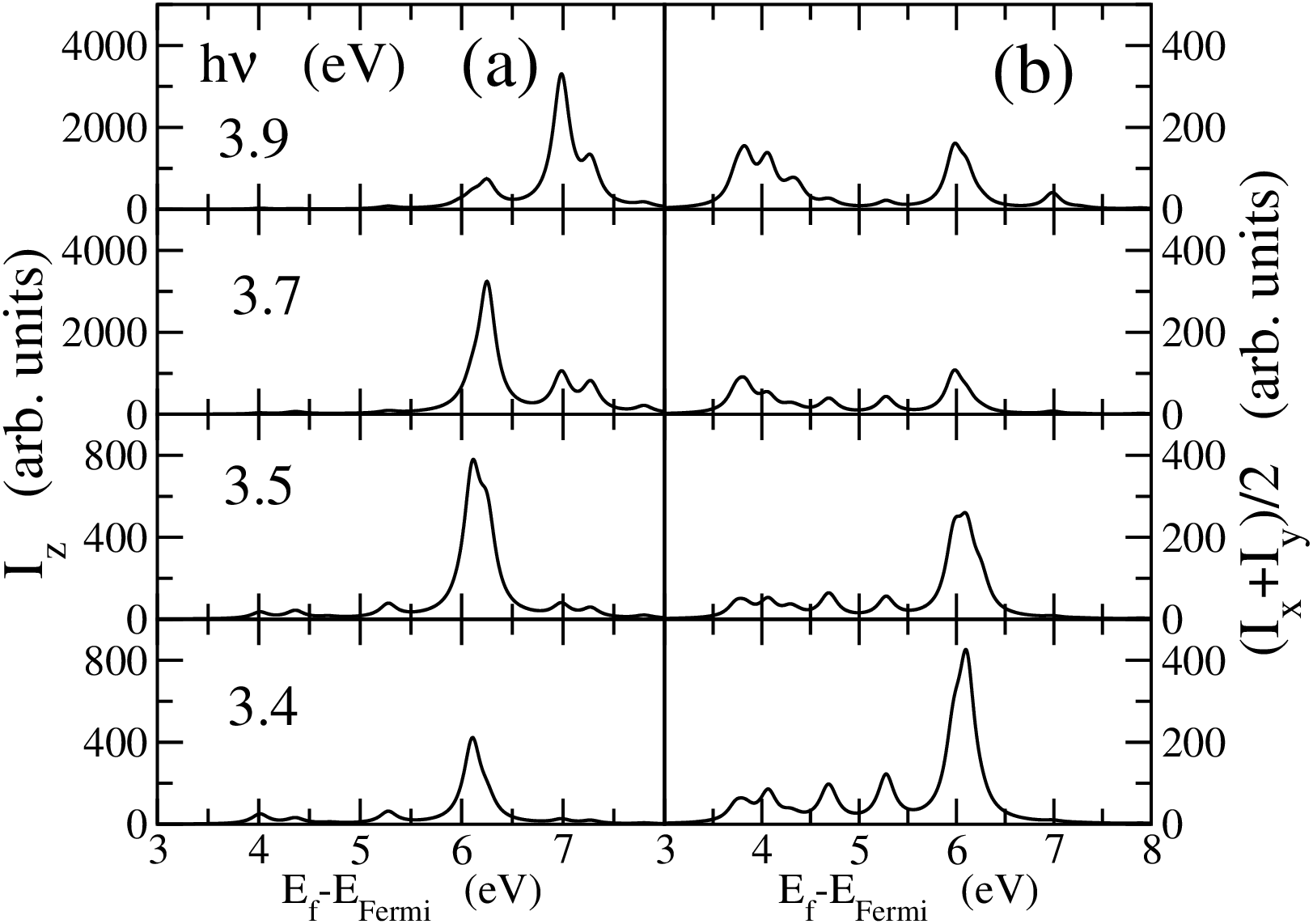}
\caption{\label{fig6} Prediction of the 2PPE spectra for the
direct mechanism according to Eq. (\protect{\ref{PPd}}) for (a)
out of plane and (b) in plane light polarizations.}
\end{figure}

There are two possible 2PPE excitation mechanisms:  direct and
indirect. In the direct process with one-color pump and probe
laser frequencies, the 2PPE intensity for $\alpha$-light
polarization is \cite{Madelung,Wolf}:
\begin{eqnarray}
&&{\rm I}^{\rm dir}_{\alpha}(i,f,\omega_L)
\sim\delta(2\hbar\omega_L-{\rm E}_f+{\rm E}_i) \nonumber \\
&& \left|\sum_{i<k<f}\frac{<f|p_{\alpha}|k><k|p_{\alpha}|i>}{{\rm
E}_k- {\rm E}_i-\hbar\omega_L+i\Gamma_k}\right|^2f_i(1-f_k)
\label{PPd}
\end{eqnarray}
where $\Gamma_k$ is the lifetime broadening of intermediate level
$k$ and a denotes the Cartesian component(x,y,or z). The sum over intermediate
states in Eq.~(\ref{PPd}) includes interference effects. The delta function may
be replaced by a Lorentzian to mimic the lifetime of the final state and the
width of the laser pulse. The result for the
$\Gamma$-point is shown in Fig. \ref{fig6} where intensity
contributions Eq.~(\ref{PPd}) are summed over all initial
states {\it i}. The calculated curves were convoluted with
a Lorentzian function ($\Gamma=0.1$ eV) to account for the
spectrometer resolution. We use intermediate and final level
lifetimes $\Gamma_k=\Gamma_f=0.1$ eV. There is one pronounced peak
at about 6.0 eV, whose intensity is resonant with the laser
frequency at about 3.6 eV. As the excitation frequency increases
the intensity transfers to higher energy final states ($\approx
7$ eV). The resonance enhancement factor is sensitive to the
choice of the damping parameters, and even for the large value (0.1
eV) used in Fig. (\ref{fig6}), the intensity of the out-of-plane
polarization varies by an order of magnitude. The resonance
effect is related to the large dipole matrix element between the
occupied surface state SS$_{\rm o}$ and the second LUMO (band 6)
separated by 3.7 eV. For the in-plane polarization there is no
strong laser frequency dependence of the 2PPE intensity, because there
is no single pair of states which dominates the sum in Eq.
(\ref{PPd}). Therefore the ratio of the out-of-plane and the
in-plane polarizations in the direct mechanism model should be
resonant with the laser frequency.

\begin{figure}
\includegraphics[height=2.347in,width=3.16in,angle=0]{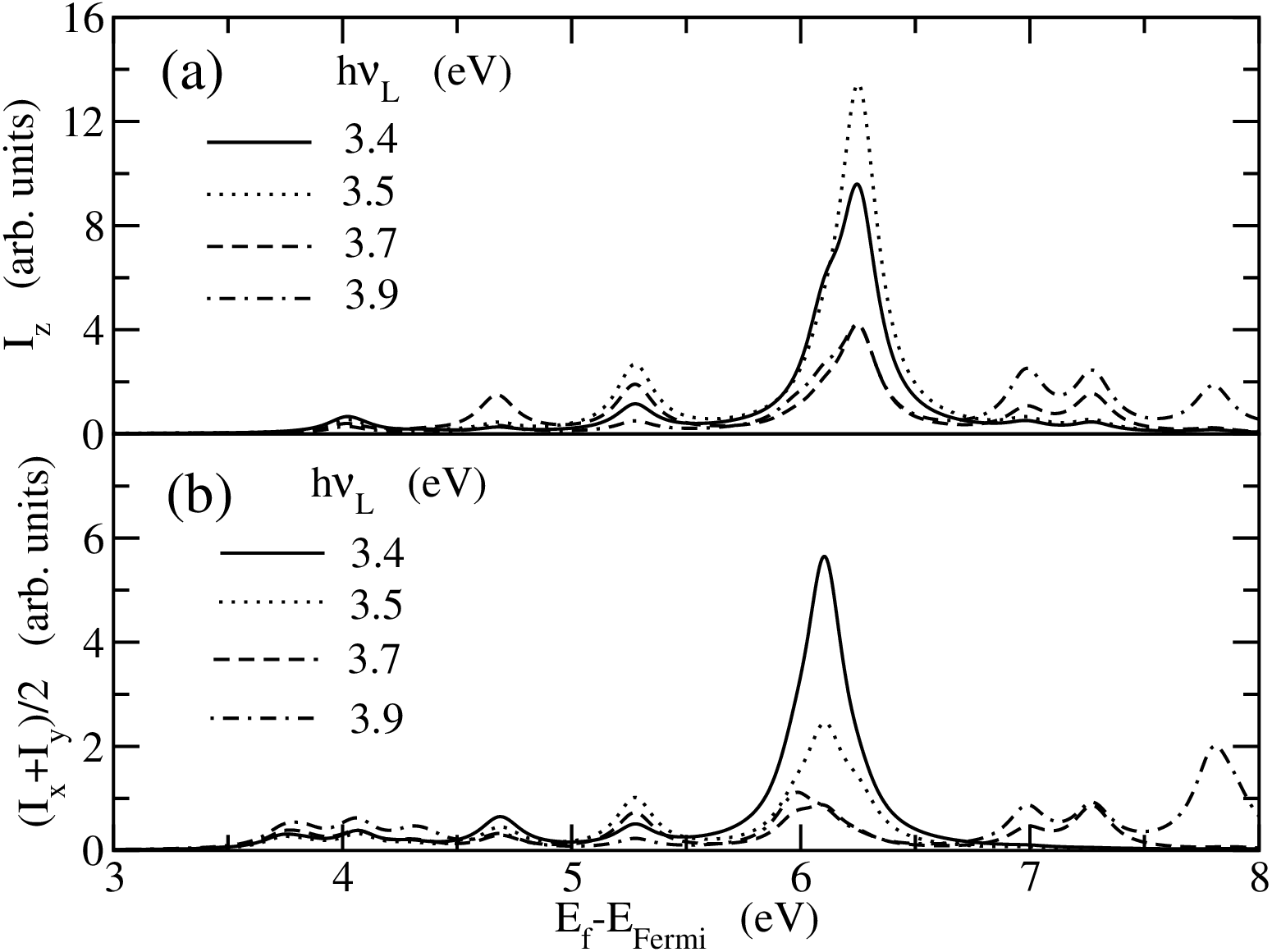}
\caption{\label{fig7} Substrate assisted 2PPE spectra ( indirect mechanism)
calculated
with Eq. (\protect{\ref{PPd2}}) for (a) out of plane and (b) in
plane light polarizations.}
\end{figure}

In the indirect mechanism the intermediate level can be populated
by some incoherent process, for example, by the tunnelling of the
photoexcited electron from the substrate to the molecule. In the
second step, a photon promotes the photoexcited electron to the
vacuum with probability given by the one photon absorption
crossection:
\begin{eqnarray}
&&{\rm I}^{\rm indir}_{\alpha}(k,f,\omega_L)
\sim\delta(\hbar\omega_L-{\rm E}_f+{\rm E}_k)
\left|<f|p_{\alpha}|k>\right|^2
\nonumber \\
&&(1-f_k)(1-f_f)f_k({\rm E_{Fermi}+\hbar\omega_L}) \label{PPd2}
\end{eqnarray}
where Fermi-Dirac factors $f$ require  that the intermediate and
final  states are  empty and the intermediate state is less than
one quantum of photon energy above the Fermi level ${\rm E_{Fermi}}$.  It is
assumed that all the intermediate states {\it k} have the equal
lifetimes and probabilities to be excited in the first step. The
result for the two polarizations and the same damping parameters
as in direct process is shown in Fig. \ref{fig7}. There are no
pronounced resonances and the ratio of intensities for the out-of-plane and the
in-plane polarizations is relatively independent of the laser
frequency. Namely, the ratios of the integrated peak intensities
at 6 eV for $\omega_L=$3.4, 3.5, 3.7, and 3.9 are 1.8, 4.6, 3.8,
and 4.0 respectively. This is consistent with the experimentally
reported value of $\approx4$ at $\omega_L=3.7$ eV \cite{Zhu}.
Comparison of our analysis with the available experimental data
\cite{Zhu} leads us to suggest that the most likely mechanism
for the two photon photoemission process is indirect. An enhancement of
2PPE intensity with the laser frequency was
observed for SCH$_3$ on Ag(111) by the Harris group
\cite{Harris}. This is may be the same effect we find here for the
direct mechanism.

\section{\label{sec5}Work function calculation}

\begin{figure}
\includegraphics[height=2.26in,width=3.19in,angle=0]{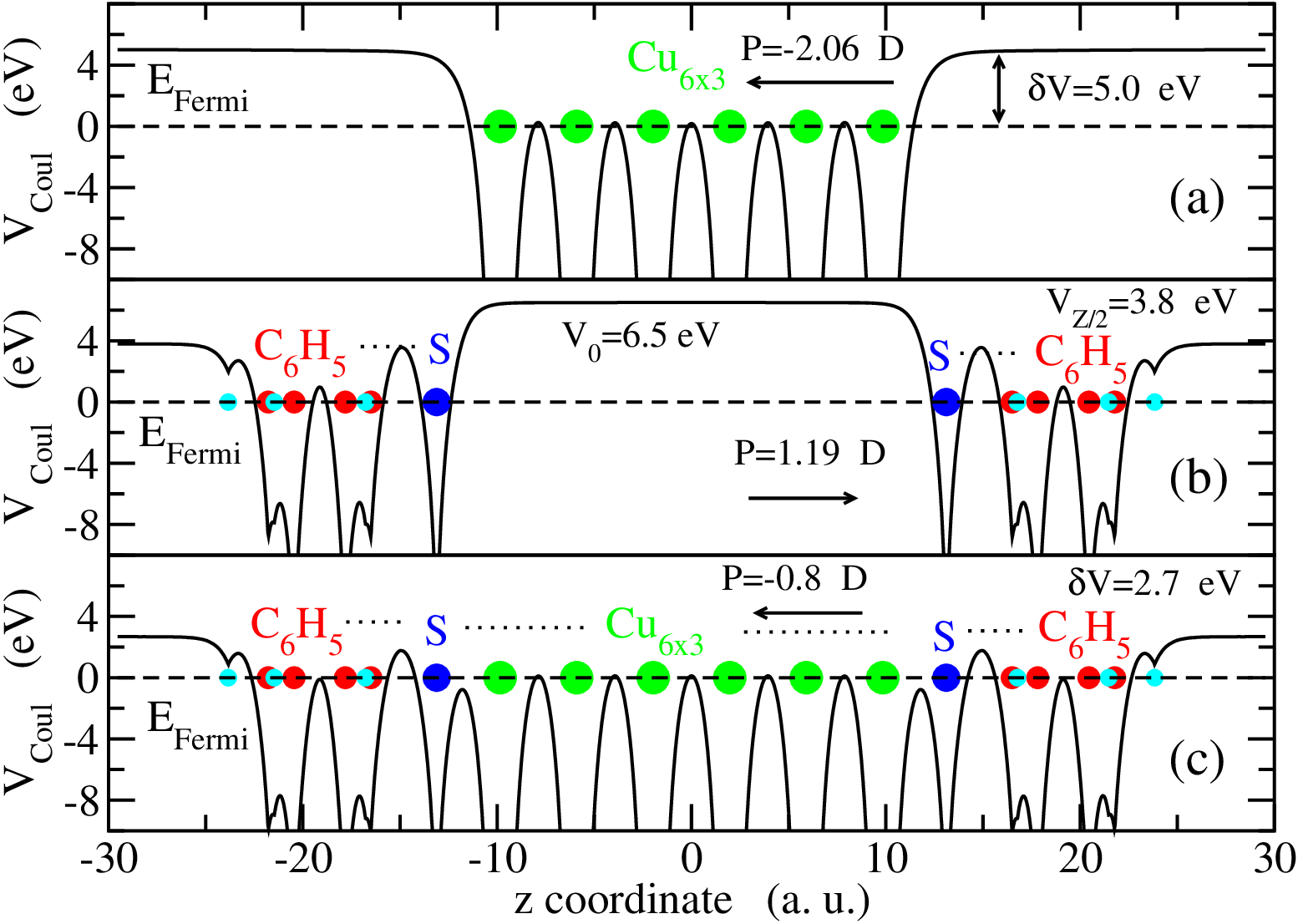}
\caption{\label{fig8} (color online). Coulomb potential averaged
in the lateral direction versus $z$-direction for (a) the clean
Cu(111), (b) the single monolayer, and (c) the SAM on Cu(111). The
dipole moments according to Eq.~(\protect{\ref{eq20}}) are shown
by the arrows.}
\end{figure}

For the work function calculations we use a six Cu layer  substrate
covered with a SAM on both sides. The inversion symmetry of the
supercell with two equivalent surfaces forces the net dipole and
the electrostatic potential drop in the vacuum region to be zero.
To separate the electrostatic and exchange correlation
contributions to the work function we calculate the averaged
surface dipole moment by integrating over the net charge density of the
half unit cell:
\begin{eqnarray}
P=\int_0^{Z/2}z[\rho(x,y,z)-\rho_{+}(x,y,z)]dxdydz, \label{eq20}
\end{eqnarray}
where $\rho$ and $\rho_+$ are valence electron and ion charge
densities respectively. The planes $z=0$ and $z=Z/2$ contain the
inversion centers of the supecell. The Coulomb potential averaged
in the lateral direction is shown on Fig. \ref{fig8} along with
that for the clean substrate and the two SAM arrays in the absence of the
substrate for the same geometry. The work function of the clean
Cu(111) is found to be 5.0 eV, in good agreement with the
experimental value of 4.9 eV. The work function of the covered
substrate is reduced to 2.7 eV compared with the experimental
value of 3.7 eV \cite{Zhu}.

\begin{figure}
\includegraphics[height=2.25in,width=3.077in,angle=0]{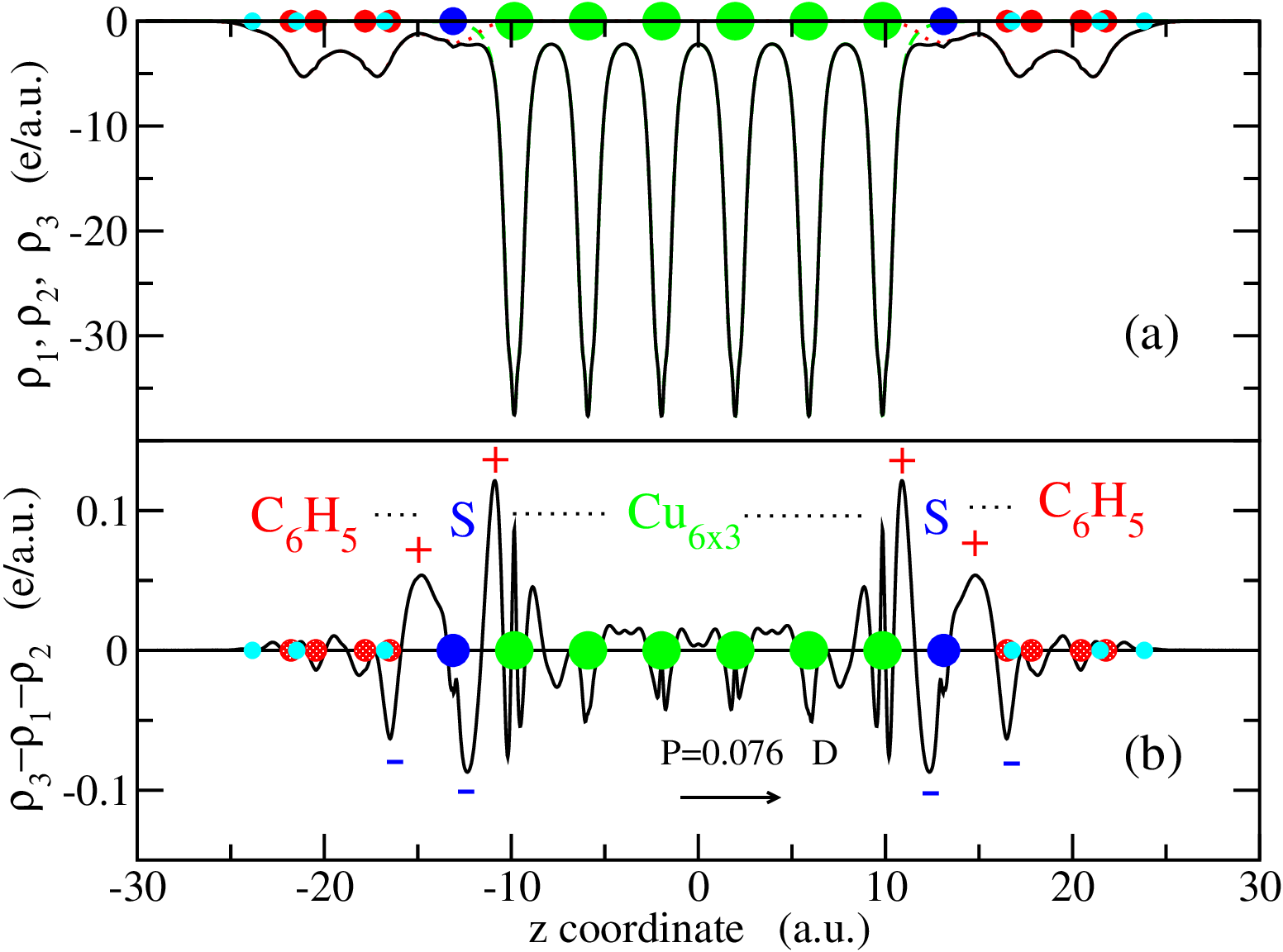}
\caption{\label{fig9}(color online). (a) The laterally average
charges densities versus $z$-direction for the clean Cu(111)
(blue) - $\rho_1$, isolated SAM array (red) - $\rho_2$ and
two-sided SAM/Cu(111) (black) - $\rho_3$. (b) The laterally
averaged charge density difference $\rho_3-\rho_1-\rho_2$ versus
$z$-direction which integrates to the induced dipole P=0.08 (D)
due to the charge transfer. }
\end{figure}

The isolated SAM layer has surface dipole $P_{\rm SAM}=1.19$
(Debye/molecule), and the expected Coulomb potential drop of $4\pi
P_{\rm SAM}/A=2.67$ eV (where $A=16.9$\AA$^2$ is the lateral area
per molecule) agrees well with the actual result $\delta
V=6.5-3.8=2.7$ eV (see Fig.~(\ref{fig8}b)). Analogously we define
the electrostatic contribution to the clean Cu surface
workfunction  to be $4\pi P_{\rm Cu}/A=4.6$ eV (see Fig.
(\ref{fig8}a)). This leaves a 0.4 eV contribution due to exchange-correlation
effects. In the case of the SAM on Cu, the net surface dipole is
approximately equal to the sum of dipoles of the clean Cu surface
and the isolated SAM array $-2.06+1.19=-0.87$ Debye. The
difference between the simple estimate and the actual result
($-0.076$ Debye) is due to the charge transfer at the interface. The
laterally averaged charge  densities are shown in Fig.
(\ref{fig9}a). The charge transfer electron density shown in Fig.
(\ref{fig9}b) is very small (less then 0.1 $e$ per molecule). We
calculate the workfunction change of the covered substrate to be
-2.3 eV, where -2.8 eV is the electrostatic and 0.5 eV is the
exchange-correlation contribution,respectively.

As we have shown the electrostatic contribution to the
workfunction change can be immediately estimated from the dipole
moment of the isolated SAM array. However, the isolated SAM dipole
moment itself is not equal to that of isolated  molecule. We have used
the molecular NRLMOL code \cite{Mark} and find the dipole moment of the
S-C$_6$H$_6$ molecule to be
 3.33 Debye. This value is threefold larger than the dipole
per molecule in the array, 1.19 Debye. We have also performed
constrained calculations by forcing occupations of two nearly
degenerate HOMO orbitals: $p_x$ (in the molecular plane (xz)) and
$p_y$ (perpendicular to the molecular plane). The unconstrained
spin restricted ( ie, pure doublet) calculation  results in 1.88
electrons in $p_x$ and 1.12 electrons in $p_y$ orbitals, with a
net dipole moment of 3.33 Debye. By forcing the $p_y$ orbital to
be fully occupied the net dipole moment drops to 1.24 Debye,very
close to the SAM value. This suggests that the charge
redistribution between the two HOMO orbitals in the SAM is the
primary effect on the net dipole moment.

\section{Discussion and Conclusions}

We have calculated the electronic structure of the phenylthiolate
(PT=S-C$_6$H$_5$) SAM on Cu(111) and Au(111) substrates. We identified
four HOMO's and three LUMO's for each system. A sulfur-type HOMO
(band 4) is the closest to the Fermi level for both substrates. It
has two competing contributions to the lateral dispersion: (1) the
direct exchange between the neighboring orbitals leads to a
negative effective mass, whereas  (2) the substrate-mediated
interaction gives a positive effective mass contribution. The
overall sign of the effective mass is changed from a negative to a
positive value in going from the Cu to the Au substrate. The most
dispersive molecular bands are of 'side carbon' character
(C$_{\rm s}$). The significant lateral dispersion of the electronic
states derived from molecular orbitals of the SAM suggests that
the charge injection probability depends on the lateral component of the
wavevector as well as the energy of the
injected electron. This implies that the conductance of N molecules in an
array is not simply N times the conductance of one, but further
theoretical and experimental work is needed to quantify the
lateral dispersion effect on transport.

We have calculated the 2PPE spectra of a PT SAM on Cu(111) using the DFT
wavefunctions. In the direct mechanism we predict a very strong
resonance of the 2PPE intensity for out-of-plane polarization.
For the in-plane polarization only small intensity variation with
laser frequency is found. For the indirect mechanism we find a
moderate intensity dependence on the excitation photon energy for
both polarizations with a nearly constant ratio of the
out-of-plane to the in-plane signals of about 3-4 , consistent
with related experimental rsults \cite{Zhu}. This fact suggests that the
indirect mechanism contributes the most intensity to the 2PPE
resonance observed in \cite{Zhu} ( designated as resonance A). The energy
positions of the observed molecular states at 6.4 eV and 3.4 eV are in
reasonable correspondence
with our calculated values of 6.0 eV and 2.8 eV,respectively.
These molecular states are mainly due to LUMOs dominated by the PT carbon atoms
adjacent to the S atoms.

Calculation of the workfunction for the PT-SAM on Cu(111) yields a
value of 2.7 eV, which may be compared with the experimental value
of 3.7 eV.  To reconcile theory and experiment one may assume a
lower coverage of the SAM in the experimental system. The larger
effective area per molecule reduces the surface dipole moment and
hence the Coulomb contribution to the workfunction change.
According to our calculations one has to assume $A\approx23.8$
\AA$^2$ to agree with experiment. This estimate is much larger
than that for the $\sqrt{3}\times\sqrt{3}R30^0$ structure, which
implies $A=16.9$ \AA$^2$. Alternatively, the interpretation of the
workfunction measurements in one photon photoemission spectra can
be complicated if there is an exit barrier for substrate electrons
to tunnel through the interface. Then the onset of the one-photon
photoemission signal may be due to the finite energy electrons
with respect to the vacuum level. This in turn would imply that
the agreement between the theory and the experiment for the
energies of the LUMO orbitals measured in 2PPE is accidental.
Indeed DFT is known to underestimate the bandgaps in
semiconductors. On the other hand the energy differences between
the empty states are more reliable in DFT and in the experiment
the energy difference between the two LUMO's is also independent
of the workfunction value. We find an electrostatic contribution
to the work function change of 2.8 eV ( associated with SAM
formation) to be the dominant effect. Therefore the dipole moment
of the molecular SAM in the absence of the metal substrate can be
used to predict the the workfunction change of the covered
substrate provided the charge transfer and correlation effects do
not contribute much to the workfunction change as in the case
studied here.

\begin{acknowledgments}
The computations were performed on the BNL galaxy cluster. We are
grateful to Mike Weinert, Ben Ocko, and X.-Y. Zhu for helpful
discussions. This work was supported in part by DOE  Grant No.\
DE-AC-02-98CH10886.
\end{acknowledgments}


\end{document}